\documentclass[aps]{revtex4}
\usepackage{amssymb,epsf}

\begin{document}

\title{Nonsingular universe in massive gravity's rainbow}
\author{Seyed Hossein Hendi$^{1,2}$\footnote{%
email address: hendi@shirazu.ac.ir}, Mehrab
Momennia$^{1}$\footnote{
email address: m.momennia@shirazu.ac.ir}, Behzad Eslam Panah$^{1}$ \footnote{%
email address: behzad.eslampanah@gmail.com} and Shahram Panahiyan$^{1,3}$ \footnote{%
email address: sh.panahiyan@gmail.com}} \affiliation{$^1$Physics
Department and Biruni Observatory, College of Sciences, Shiraz
University, Shiraz 71454, Iran\\
$^2$ Research Institute for Astronomy and Astrophysics of Maragha (RIAAM),
Maragha, Iran \\
$^{3}$ Physics Department, Shahid Beheshti University, Tehran 19839, Iran}

\begin{abstract}
One of the fundamental open questions in cosmology is whether we can regard
the universe evolution without singularity like a Big Bang or a Big Rip.
This challenging subject stimulates one to regard a nonsingular universe in
the far past with an arbitrarily large vacuum energy. Considering the high
energy regime in the cosmic history, it is believed that Einstein gravity
should be corrected to an effective energy dependent theory which could be
acquired by gravity's rainbow. On the other hand, employing massive gravity
provided us with solutions to some of the long standing fundamental problems
of cosmology such as cosmological constant problem and self acceleration of
the universe. Considering these aspects of gravity's rainbow and massive
gravity, in this paper, we initiate studying FRW cosmology in the massive
gravity's rainbow formalism. At first, we show that although massive gravity
modifies the FRW cosmology, but it does not itself remove the big bang
singularity. Then, we generalize the massive gravity to the case of energy
dependent spacetime and find that massive gravity's rainbow can remove the
early universe singularity. We bring together all the essential conditions
for having a nonsingular universe and the effects of both gravity's rainbow
and massive gravity generalizations on such criteria are determined.
\end{abstract}

\maketitle

\section{Introduction}

The possible existence of big bang singularity at the early universe and its
challenging physical properties (at still earlier epochs near the Planck
energy, $E_{p}$) urge one to look for a quantum theory of gravity \cite%
{NovelloB,AshtekarS,Pinto-NetoF,GarayMM,Calcagni,Brandenberger,BattefeldP}.
It is believed that the quantum gravity admits a semi-classical regime \cite%
{Smolin,GarattiniM,LingMPLA} in which the quantum corrections could be
viewed as dependency of the spacetime metric on the energy of particles
probing it. Thus, we should regard essentially an upper bound on the energy
scale of such particles which is Planck energy. The possible connection
between the energy of probing particles and the energy-dependent spacetime
is proposed in the context of gravity's rainbow. In the gravity's rainbow
framework, it is considered that particles with different energies are
affected differently by the structure of spacetime, depending on their
wavelengths \cite{Magueijo}.

one of the basic elements for building gravity's rainbow theory is modified
energy-momentum dispersion relation. This modification can be arisen from
applying nonlinear Lorentz transformations with the following explicit
result
\begin{equation}
E^{2}f(\varepsilon )^{2}-p^{2}g(\varepsilon )^{2}=m_{0}^{2},  \label{DeforEM}
\end{equation}%
where the energy ratio is $\varepsilon =E/E_{p}$ in which $E$ is
the energy of test particle. Besides, both $f(\varepsilon )$ and
$g(\varepsilon )$ are rainbow functions, and $m_{0}$ is the mass
of the test particle. It is also
notable that in the limit of ${\lim_{\varepsilon \rightarrow 0} }%
f(\varepsilon )=1$ and ${\lim_{\varepsilon \rightarrow 0} }g(\varepsilon )=1$%
, the standard energy-momentum dispersion relation is recovered.
Such modification in energy-momentum dispersion relation has been
supported in the context of several topics, such as discrete
spacetime \cite{Hooft}, spacetime foam \cite{Amelino-CameliaEMNS},
spin-network in loop quantum gravity (LQG) \cite{Gambini}, ghost
condensation \cite{FaizalJPA} and non-commutative geometry
\cite{Carroll,FaizalMPLA}. Moreover, the observational data of the
Pierre Auger Collaboration and the High Resolution Fly's Eye
experiment \cite{Abraham} have confirmed the requirement for such
modification and also the necessity of an upper limit on the
energy of cosmic rays.

Another cornerstone of considering the gravity's rainbow is based
on the generalization of doubly special relativity
\cite{MagueijoPRD} in curved spacetime \cite{Magueijo}. The doubly
special relativity is a generalization of the special relativity
which considers an upper limit for energies that a particle can
acquire. Particularly, its generalization to non-flat spacetimes
is the so-called doubly general relativity. Besides, using the
gravity dictionary, one may rename the doubly general relativity
to gravity's rainbow. Among the interesting achievements of the
gravity's rainbow, one can regard the UV completion of general relativity \cite%
{Magueijo}, existence of remnants for black holes after evaporation \cite%
{AliNPB,AliPRD2014}, admitting usual uncertainty principle \cite%
{LingMPLA,LingCQG} and providing solutions for information paradox \cite%
{AliEPL,Gim2015}. Moreover, the gravity's rainbow has been
employed to investigate the thermodynamical properties of black holes \cite%
{HendiF,HendiEParXiv,Hendi,HendiPEMEPJC,HendiFEP} and the
structure of neutron stars \cite{rainbowTOV}. In the context of
cosmology, it was shown that employing this formalism will provide
the possibility of removing the big bang singularity
\cite{cosmosAwad,cosmosLing,cosmosSantos,Hendicosmology} and the
big bounce of a cyclic universe \cite{LingWu}. The mentioned
subjects motivate us to consider the gravity's rainbow in the
context of modified Einstein theory. It is worth mentioning that
through several works, it was shown that by using the nonlinear
electrodynamics, one can remove the possibility of existence of
singularity in early universe
\cite{nonsing1,nonsing2,nonsing3,nonsing4}. Here, our idea is to
employ the approach of gravity's rainbow alongside of massive
gravity instead of the nonlinear electromagnetic field for
removing the big bang singularity.

Einstein theory of gravity incorporates massless spin$-2$
gravitons as intermediate couriers of gravitational interactions \cite%
{Gupta,Weinberg,Deser,BoulwareDI}. But studies that are conducted in the
context of brane-world gravity (especially regarding hierarchy problem)
emphasize on the existence of massive gravitons \cite%
{massiveg,DvaliGP,DvaliGPI,DvaliG}. Therefore, it is logical to generalize
Einstein gravity to a new theory with massive gravitons. This generalization
can improve our insight concerning the gravitational concepts. In this
regard, the first attempts were done by Fierz and Pauli \cite{Fierz}. This
theory of massive gravity suffers the existence of vDVZ (van
Dam-Veltman-Zakharov) discontinuity which indicates that for the limit of
vanishing mass, the propagators of massive and massless theories are not
consistent \cite{van Dam,Zakharov,DeserW}. One of the solutions of this
problem is Vainshtein mechanism which requires a system to be considered in
a nonlinear framework \cite{Vainshtein}. But the generalization to
nonlinearity leads to the presence of extra degree of freedom which provides
a ghost instability \cite{Boulware}. To overcome this problem, several
theories of the massive gravity have been proposed, in which among them one
can name new massive gravity \cite{NEW}, bi-gravity \cite{bi} and dRGT (de
Rham, Gabadadze and Tolley) theory \cite{de RhamGT,de RhamGTI,Hinterbichler}%
. The dGRT massive gravity is a ghost free theory which employs a reference
metric to build massive terms \cite{de RhamGT,de
RhamGTI,Hinterbichler,HassanRI,HassanRS}. The modification in the reference
metric's ansatz leads to different theories of massive gravity which are
dGRT-like \cite{review}. One of these modifications with specific interest
in gauge/gravity duality was done by Vegh where the graviton plays the role
of lattice \cite{Vegh}. This theory of massive gravity has been employed to
conduct studies regarding super massive objects such as black holes \cite%
{HendiEParXivI,HendiEParXiv,HendiEPJHEPI,HendiPEJ,HendiEPJHEPII}. It was
shown that the presence of massive gravitons has important contributions
into geometrical, thermodynamical and structure of these super massive
objects. Also, the existence of massive gravitons can generate a new range
of phase transitions for topological black holes \cite{PRDr}. From
cosmological perspective, massive gravity has also a long history. Its
effects on the existence of flat and open FRW universes were studied before
\cite{massivecosmologyA1,massivecosmologyA2}. In addition, the ghostlike
(in)stabilities of massive cosmology have been investigated in literature
\cite{KhosraviNK,Felice,bi}. Also, it was pointed out that the massive
theory can be used to explain the cosmological constant problem \cite%
{DvaliGS,DvaliHK} and provides an interesting basis for self-acceleration of
the universe without introducing the cosmological constant \cite%
{Deffayet,DeffayetDG}. In other words, it was shown that the massive
graviton term in cosmological solutions can be equivalent to a cosmological
constant \cite{massivecosmologyA2,GratiaHW,Kobayashi}. Furthermore, the
existence of massive gravitons provides extra polarization for gravitational
waves, and affects the propagation's speed of gravitational waves \cite{GW1}
and production of gravitational waves during inflation \cite{GW2,GW3}.
Therefore, it is reasonable to take the massive gravity into account for the
cosmological systems. It is expected that investigation of the early
universe in the context of massive gravity opens a window to discuss
cosmological implications at high energy regime.

In $2016$, for the first time, gravitational waves produced by a
binary black hole merger were detected by the LIGO collaboration
by using laser interferometer method \cite{LIGO1,LIGO2}. This
confirmed the validation of the general relativity with two of its
predictions, known as the existence of black holes and
gravitational waves. In addition, as it was pointed out in Ref.
\cite{discriminating}, realization of a gravitational wave
astronomy provides us with the possibility of discriminating among
general relativity and other gravity theories. In other words, by
improving the results of LIGO and Virgo scientific collaborations,
we are able to determine the validation of Einstein gravity
extensions. Here, two of such extensions are considered alongside
each other; gravity's rainbow and massive gravity. In this paper,
we take into account the modified FRW universe in the massive
gravity's rainbow. The main motivations for such consideration is
given as follows; first of all, the physical properties of early
universe dictate the necessity of regarding semi-quantum
corrections which could be achieved by considering geometry of the
spacetime as an energy dependent one. To do so, we employ
gravity's rainbow. This consideration has specific contributions
such as absence of the big bang in the history of universe. Since
the evolution of universe is governed by gravity, it is crucial to
examine and understand the effects of massive gravitons on
different stages of the universe's evolution such as big bang,
inflation and etc. Especially, it is important to understand the
effects of massive gravitons on gravitational waves which were
produced during inflation. Therefore, in this paper, we intend to
provide preliminaries for such studies and investigate massive
gravity, too. In essence, when we are dealing with a physical
system, we should impose maximum number of generalizations, in
order to have more general results and reliable predications. This
is another motivation why we are considering such set up for our
gravitational system.

The outline of the paper is as follows. In the next section, we
study the Einstein-massive gravity and show that although
Friedmann equations are modified in the presence of massive
gravity, the singular big bang remains unsolved. After that, we
modify FRW universe in Einstein-massive gravity's rainbow, and
then we show that there is no big bang singularity. We investigate
nonsingular rainbow universe by considering three cases of rainbow
functions. We finish our paper with some concluding
remarks.\newline

\section{Modified FRW cosmology in the massive gravity}

Here, we are interested in the structure of FRW universe in the context of
massive gravity. We consider the Lagrangian of Einstein-massive gravity with
a matter field as
\begin{equation}
\mathcal{L}_{E}=\mathcal{R}+m^{2}\sum_{i=1}^{4}c_{i}\mathcal{U}_{i}(g,f)+%
\mathcal{L}_{m},  \label{LE}
\end{equation}%
where $\mathcal{R}$ and $\mathcal{L}_{m}$ are the Lagrangians of Einstein
gravity and matter field, respectively, and also $m$ is the massive
parameter. The second term in the Lagrangian of system produces massive
terms in which $f$ is a fixed symmetric tensor, $c_{i}$'s are some
constants, and $\mathcal{U}_{i}$'s are symmetric polynomials of the
eigenvalues of matrix $K_{\nu }^{\mu }=\sqrt{g^{\mu \alpha }f_{\alpha \nu }}$
\begin{eqnarray*}
\mathcal{U}_{1} &=&\left[ \mathcal{K}\right] , \\
\mathcal{U}_{2}&=&\left[ \mathcal{K}\right] ^{2}-\left[ \mathcal{K}^{2}%
\right] , \\
\mathcal{U}_{3} &=&\left[ \mathcal{K}\right] ^{3}-3\left[ \mathcal{K}\right] %
\left[ \mathcal{K}^{2}\right] +2\left[ \mathcal{K}^{3}\right] , \\
\mathcal{U}_{4} &=&\left[ \mathcal{K}\right] ^{4}-6\left[ \mathcal{K}^{2}%
\right] \left[ \mathcal{K}\right] ^{2}+8\left[ \mathcal{K}^{3}\right] \left[
\mathcal{K}\right] +3\left[ \mathcal{K}^{2}\right] ^{2}-6\left[ \mathcal{K}%
^{4}\right] .
\end{eqnarray*}

Using the variational principle, one can find following field equations
\begin{equation}
R_{\mu \nu }-\frac{1}{2}\mathcal{R}g_{\mu \nu }+m^{2}\chi _{\mu \nu }=8\pi
GT_{\mu \nu },  \label{FieldeqE}
\end{equation}%
in which $\chi _{\mu \nu }$ is the massive term with the following form%
\begin{eqnarray}
&&\chi _{\mu \nu } =-\frac{c_{1}}{2}\left( \mathcal{U}_{1}g_{\mu \nu }-%
\mathcal{K}_{\mu \nu }\right) -\frac{c_{2}}{2}\left( \mathcal{U}_{2}g_{\mu
\nu }-2\mathcal{U}_{1}\mathcal{K}_{\mu \nu }+2\mathcal{K}_{\mu \nu
}^{2}\right)  \nonumber \\
&&-\frac{c_{3}}{2}(\mathcal{U}_{3}g_{\mu \nu }-3\mathcal{U}_{2}\mathcal{K}%
_{\mu \nu }+6\mathcal{U}_{1}\mathcal{K}_{\mu \nu }^{2}-6\mathcal{K}_{\mu \nu
}^{3})  \nonumber \\
&&-\frac{c_{4}}{2}(\mathcal{U}_{4}g_{\mu \nu }-4\mathcal{U}_{3}\mathcal{K}%
_{\mu \nu }+12\mathcal{U}_{2}\mathcal{K}_{\mu \nu }^{2}-24\mathcal{U}_{1}%
\mathcal{K}_{\mu \nu }^{3}+24\mathcal{K}_{\mu \nu }^{4}).
\end{eqnarray}

The $4-$dimensional FRW metric with flat horizon ($k=0$) is given by
\begin{equation}
ds^{2}=-dt^{2}+R(t)^{2}dx_{i}^{2},~~\ \ \ \ \ \ i=1,2,3,  \label{metric}
\end{equation}%
where $R(t)$ is the scale factor. The energy-momentum tensor of our
cosmological system is constructed based on perfect fluid as
\begin{equation}
T_{\mu \nu }=\rho (t)u_{\mu }u_{\nu }+P(t)(g_{\mu \nu }+u_{\mu }u_{\nu }),
\label{EMTensor}
\end{equation}%
where $\rho (t)$ is the energy density and $P(t)$ is the pressure of perfect
fluids in which we will use $\rho $\ and $P$\ for simplicity. $u_{\mu }$ is
four vector velocity which is defined as
\begin{equation}
u_{\mu }=(1,0,0,0),
\end{equation}%
where $u_{\mu }$ satisfies the following restriction
\begin{equation}
g^{\mu \nu }u_{\mu }u_{\nu }=-1.
\end{equation}

The massive terms are constructed by considering a reference
metric. Here, we employ the following ansatz for the reference
metric \cite{CaiMassive}
\begin{equation}
f_{\mu \nu }=diag(0,0,c^{2}h_{ij}),  \label{f11}
\end{equation}%
in which $c$ is a positive constant. It is worthwhile to mention that this
reference metric preserves general covariance in temporal and radial
coordinates but not in the transverse spatial coordinates \cite{Vegh}, and
therefore, the massive terms will have a Lorentz breaking property. In other
words, we relax the Lorentz invariance property of massive gravity \cite%
{Yagi}, which may be a necessary requirement for quantum description of
gravity \cite{Hooft,Amelino-CameliaEMNS,quantum1,quantum2,quantum3,Amelino}.
In addition, we are not concerned to encounter singular properties of the
reference metric (which is not as a physical metric, but as a tool), since
we use the line element $g_{\mu \nu}$ for raising and lowering indices.
Using this metric ansatz (\ref{f11}), $\mathcal{U}_{i}$'s are constructed in
the following forms \cite{CaiMassive}
\begin{equation}
\mathcal{U}_{1}=\frac{2c}{r},\;\;\mathcal{U}_{2}=\frac{2c^{2}}{r^{2}},~~%
\mathcal{U}_{3}=\mathcal{U}_{4}=0,  \label{U}
\end{equation}%
in which $r$ is the co-moving coordinate. Considering the metric (\ref%
{metric}) and field equation (\ref{FieldeqE}), one can find the modified FRW
equations in the massive gravity as
\begin{equation}
3H^{2}+\frac{cm^{2}}{r^{2}}\left( c_{1}r+cc_{2}\right) =8\pi G\rho ,
\label{FEQE1}
\end{equation}%
\begin{equation}
\dot{H}=-4\pi G(\rho +P),  \label{FEQE3}
\end{equation}%
where $H=\dot{R}/R$ is the Hubble parameter and we used the notation $\dot{A}%
=\frac{dA}{dt}$. It is notable that, in the absence of massive parameter ($%
m=0$), the Friedmann equations in Einstein-massive gravity (\ref{FEQE1} and %
\ref{FEQE3}) reduce to the ones in Einstein gravity. Now, we consider the
conservation equation of energy-momentum tensor as
\begin{equation}
\nabla _{\mu }T_{\nu }^{\mu }=\partial _{\mu }T_{\nu }^{\mu }-\Gamma _{\mu
\nu }^{\lambda }T_{\lambda }^{\mu }+\Gamma _{\mu \lambda }^{\mu }T_{\nu
}^{\lambda }=0,
\end{equation}%
where after some calculations, one finds
\begin{equation}
\dot{\rho}+3H(\rho +P)=0.  \label{ConsEqE}
\end{equation}

Here, we are interested in a large range of ultra relativistic particles
which are in thermal equilibrium with a typical or an average energy $%
\epsilon \sim T$. Using the concept of continuity equation, one can find
following the first law of thermodynamics in standard cosmology
\begin{equation}
d(\rho V)=-PdV,  \label{PV}
\end{equation}%
where the volume, $V$, is given by $V=[R(t)]^{3}$. It is worthwhile to
mention that regarding the integrability condition ($\frac{\partial ^{2}S}{%
\partial V\partial P}=\frac{\partial ^{2}S}{\partial P\partial V}$) \cite%
{Kolb} with the first law of thermodynamics (\ref{PV}) will lead to a
constant entropy in the following form \cite{Weinbergbook}
\begin{equation}
S=\frac{V(\rho +P)}{T}=const.
\end{equation}

In order to obtain the properties of FRW cosmology, it is necessary to
consider an equation of state (EoS). Here, EoS is given by
\begin{equation}
P=(\gamma -1)\rho,  \label{EoS}
\end{equation}
which leads to a singular FRW spacetime in standard cosmology at $t=0$ ($%
\gamma$ known as EoS parameter which is $4/3$ for the radiation dominated
era). The average energy, $\epsilon$, is obtained as
\begin{equation}
\epsilon \sim T=c^{\prime }\gamma \rho V,  \label{AV}
\end{equation}%
where $c^{\prime }$. is a constant which is equal to $1/S$. Substituting the
EoS (\ref{EoS}) in the conservation equation (\ref{ConsEqE}), one can find
\begin{equation}
\frac{d\rho }{d\ln (R)}=-3\gamma \rho ,
\end{equation}%
which leads to energy density as $\rho =R^{-3\gamma }$. It is a matter of
calculation to find average energy as%
\begin{equation}
\epsilon =c^{\prime }\gamma \rho ^{\frac{\gamma -1}{\gamma }},
\label{averageEM}
\end{equation}
where the relation between the average energy and density is governed by EoS
parameter, $\gamma$.

Now, we are in a position to investigate the condition regarding singular
and nonsingular universe. To do so, we are going to follow the same analysis
that has been used in Refs. \cite{cosmosAwad,Hendicosmology}. Using the
obtained modified Friedmann equation of massive gravity (Eq. (\ref{FEQE1}))
with conservation equation (Eq. (\ref{ConsEqE})) and EoS (Eq. (\ref{EoS})),
we obtain
\begin{equation}
\dot{\rho}=\pm \gamma \rho \sqrt{24\pi G\rho -\frac{3\mathcal{C}m^{2}}{r^{2}}%
}.  \label{ConsVSrho}
\end{equation}%
where $\mathcal{C}=c\left(c_{1}r+cc_{2}\right)$. The time for reaching a
potential singularity could be obtained by integrating Eq. (\ref{ConsVSrho})
(starting from an initial finite density $\rho ^{\ast}$ to an infinite one).
Therefore, the integration is given by
\begin{equation}
t=\pm \frac{1}{\gamma }\int_{\rho ^{\ast }}^{\infty }\frac{1}{\rho }\left[
24\pi G\rho -\frac{3\mathcal{C}m^{2}}{r^{2}}\right] ^{-\frac{1}{2}}d\rho ,
\label{TM}
\end{equation}%
and it is a matter of calculation to find
\begin{equation}
t=\pm \left. \frac{2r}{\gamma m\sqrt{3\mathcal{C}}}\arctan \left( \sqrt{%
\frac{8\pi G\rho r^{2}}{\mathcal{C}m^{2}}-1}\right) \right\vert _{\rho
^{\ast }}^{\infty },
\end{equation}
in which by substituting the bounds, one can find
\begin{equation}
t=\pm \frac{2r}{\gamma m\sqrt{-3\mathcal{C}}}\sinh ^{-1}\left( \frac{m}{2r}%
\sqrt{-\frac{\mathcal{C}}{2\pi G\rho ^{\ast }}}\right) ,  \label{eq24}
\end{equation}%
where $\mathcal{C}$ should be negative ($\mathcal{C}<0$). Considering
negative $C$, it is possible to obtain a limit on massive coefficients.
Regarding the fact that the co-moving coordinate is zero at the big bang,
one can conclude that whether $c$ and $c_{2}$, both must be negative or only
$c_{2}$ itself must be negative while $c$ is positive. As we drift away from
big bang, the effects of $c_{1}r$ term start to grow, so we get another set
of condition indicating that $c$ and $c_{2}$ should be negative and $c_{1}$
must be positive, so Eq. (\ref{eq24}) has positive term for its square root
function. Another possible case for having real function in Eq. (\ref{eq24})
is by setting $c$ as positive constant while $c_{1}$ and $c_{2}$ are
negative. These two cases enable us to put limit on the signs that massive
coefficients could acquire. Regarding Eq. (\ref{eq24}), one can conclude
that the time is finite as density of universe goes from an initial density $%
\rho ^{\ast }$\ (the present time) to infinity (the big bang time). This
means that the time to reach the potential singularity is finite, so we have
the big bang singularity.\newline

\section{Modified FRW Rainbow Cosmology in Einstein-Massive Gravity's Rainbow%
}

In this section, we are going to modify usual FRW universe by considering
Einstein-massive gravity's rainbow and study its effect as a semi-classical
approach in the early universe. In other words, since dRGT massive gravity
is a low energy effective theory, one may use its modifications to address
issues such as the big bang singularity. In addition, one can keep the
gravitational structure of field equations and instead consider an
energy-dependent ansatz for the metric as a stand-in for any high energy or
quantum effects. In this prescription, we focus on the energy dependent
nature of spacetime, which is affected by the energy of moving particles. As
a starting point, we consider the following $4$-dimensional metric
\begin{equation}
ds^{2}=-\frac{dt^{2}}{k(\varepsilon )^{2}}+\frac{R(t)^{2}}{g(\varepsilon
)^{2}}dx_{i}^{2},~~\ \ \ \ \ \ i=1,2,3,  \label{metric1}
\end{equation}
and here, we define $u_{\mu }$ as
\begin{equation}
u_{\mu }=(k(\varepsilon )^{-1},0,0,0),  \label{unit}
\end{equation}
in which
\begin{equation}
g^{\mu \nu }u_{\mu }u_{\nu }=-1.
\end{equation}

We obtain the conservation equation in the following form
\begin{equation}
\dot{\rho}+3\left( H-\frac{\dot{g}(\varepsilon )}{g(\varepsilon )}\right)
(\rho +P)=0.  \label{ConsMR}
\end{equation}

One can use the same procedure, like previous section, to show that the
average energy has the following form
\begin{equation}
\epsilon =c^{\prime }\gamma \rho ^{\frac{\gamma -1}{\gamma }}.
\label{averageEMR}
\end{equation}

It is notable that, in Eq. (\ref{averageEMR}) we used $V=(R(t)/g(%
\varepsilon))^3$, and interestingly, Eq. (\ref{averageEMR}) (such as Eq. (%
\ref{averageEM})) does not depend on the rainbow functions.

Considering Eq. (\ref{unit}) and using the metric (\ref{metric1}) with field
equation (\ref{FieldeqE}), one can show that the modified FRW equations in
Einstein-massive gravity's rainbow are
\begin{equation}
\left( H-\frac{\dot{g}(\varepsilon )}{g(\varepsilon )}\right) ^{2}+\frac{%
\mathcal{C}m^{2}}{3r^{2}k(\varepsilon )^{2}}=\frac{8\pi G\rho }{%
3k(\varepsilon )^{2}},  \label{FE1}
\end{equation}%
\begin{equation}
\dot{H}+\frac{\dot{g}(\varepsilon )^{2}}{g(\varepsilon )^{2}}-\frac{\ddot {g}%
(\varepsilon )}{g(\varepsilon )}+\frac{\dot{k}(\varepsilon )}{k(\varepsilon )%
}\left( H-\frac{\dot{g}(\varepsilon )}{g(\varepsilon )}\right) =-\frac{4\pi
G(\rho +P)}{k(\varepsilon )^{2}},  \label{FE2}
\end{equation}%
in which Eqs. (\ref{FE1}) and (\ref{FE2}) turn into Eqs. (\ref{FEQE1}) and (%
\ref{FEQE3}) for $g(\varepsilon )=k(\varepsilon )=1$, respectively. On the
other hand, they reduce to the FRW equations in Einstein gravity's rainbow
\cite{cosmosAwad} in the absence of massive parameter ($m=0$).\newline

\section{When a nonsingular rainbow universe is possible?}

In this section, we are going to show that how a nonsingular rainbow
universe is possible. Here, we want to discuss this possibility in different
cases.

\subsection{Case 1: the constants are independent from energy}

In this case, we consider all the constants as energy independent ones.
Substituting the modified Friedmann equation of massive gravity's rainbow (%
\ref{FE1}) in the conservation equation (\ref{ConsMR}) and using the EoS, we
obtain
\begin{equation}
\dot{\rho}=\pm \frac{\gamma \rho }{k(\varepsilon )}\sqrt{24\pi G\rho -\frac{3%
\mathcal{C}m^{2}}{r^{2}}}.  \label{ConsVSfg}
\end{equation}

Now, we are in a position to study the resolution of finite-time
singularities. This is done by showing that the existence of an upper bound
for the density $\rho $ is reached at an infinite time. In other words,
there exists a divergence point for density which is acquired in an infinite
time. Therefore, it is not a physical singularity.

Considering Eq. (\ref{averageEMR}), it is possible to write $k$ as a
function of $\rho $ instead of $\varepsilon$ in Eq. (\ref{ConsVSfg}). Now,
it is a matter of calculation to show that all finite-time singularities
(including big bang singularity) are removed if $k$ grows asymptotically as $%
\rho ^{1/2}$, or faster such as $k\sim \rho ^{s}$ where $s\geq 1/2$. In this
case, one can calculate the time for reaching a potential singularity by
integrating Eq. (\ref{ConsVSfg}) (starting from a finite density $%
\rho^{\ast} $ to an infinite one) which leads to
\begin{equation}
t=\pm \frac{1}{\gamma }\int_{\rho ^{\ast }}^{\infty }\rho ^{s-1}\left[ 24\pi
G\rho -\frac{3\mathcal{C}m^{2}}{r^{2}}\right] ^{-\frac{1}{2}}d\rho .
\label{IntChoice}
\end{equation}

After some calculations, we obtain
\begin{eqnarray}
t &=&\pm \left. \frac{\left( \frac{\mathcal{C}m^{2}}{Gr^{2}}\right) ^{s-1}%
\sqrt{24\pi G\rho -\frac{3\mathcal{C}m^{2}}{r^{2}}}{}\mathcal{W}_{1}}{%
3(2)^{3s-1}\pi ^{s}\gamma G}\right\vert _{\rho ^{\ast }}^{\infty },
\nonumber \\
&=&\infty ,~~s\geq \frac{1}{2},  \label{TimeChoice}
\end{eqnarray}%
in which
\begin{equation}
\mathcal{W}_{1}=_{2}\mathcal{F}_{1}\left( \left[ \frac{1}{2},1-s\right] ,%
\left[ \frac{3}{2}\right] ,1-\frac{8\pi G\rho r^{2}}{\mathcal{C}m^{2}}%
\right) ,
\end{equation}%
where $_{2}\mathcal{F}_{1}$ is a hypergeometric function and $\mathcal{C}<0$%
. Considering (\ref{TimeChoice}), we conclude that the time to reach the
potential singularity is infinite for $s\geq 1/2$, so it is not a
finite-time singularity, i.e., not physical. Evidently, the energy function,
$k(\varepsilon )$, plays a crucial role to remove the big bang singularity.

\subsection{Case 2: energy dependent constants}

In this case, we consider the two constants $G$ and $m$ as functions of
energy ($G(\varepsilon )$ and $m(\varepsilon )$). With this assumption, Eq. (%
\ref{ConsVSfg}) becomes
\begin{equation}
\dot{\rho}=\pm \frac{\gamma \rho }{k(\varepsilon )}\sqrt{24\pi G(\varepsilon
)\rho -\frac{3\mathcal{C}m(\varepsilon )^{2}}{r^{2}}}.  \label{ConsVSkGm}
\end{equation}

According to Eq. (\ref{averageEMR}), we can write $G$ and $m$ as functions
of $\rho $ instead of $\varepsilon $. Following the steps of previous
section, the time for reaching a potential singularity by integrating Eq. (%
\ref{ConsVSkGm}) is obtained as
\begin{eqnarray}
t &=&\pm \frac{1}{\gamma }\int_{\rho ^{\ast }}^{\infty }\rho ^{s-1}\left[
24\pi \rho G(\rho )-\frac{3\mathcal{C}m(\rho )^{2}}{r^{2}}\right] ^{-\frac{1%
}{2}}d\rho  \nonumber \\
&=&\pm \frac{1}{\gamma }\int_{\rho ^{\ast }}^{\infty }\rho ^{s-1}\left[
24\pi \rho ^{a+1}-\frac{3\mathcal{C}\rho ^{2b}}{r^{2}}\right] ^{-\frac{1}{2}%
}d\rho ,  \label{IntMRmG}
\end{eqnarray}%
where in the above equation we considered $G(\rho )=\rho ^{a}$ and $m(\rho
)=\rho ^{b}$. One may note that it is not possible to compute this
integration analytically. Indeed, we can solve this integration without any
bounds, but it is not possible to find the asymptotic behavior of solution
because we need to know whether $a+1$ is larger (smaller) than $2b$, or they
are equal. So, we have to restrict ourselves to the special case, $G(\rho
)=m(\rho )^{2}=\rho ^{2b}$. After some calculations, we obtain
\begin{eqnarray}
t &=&\pm \frac{1}{\gamma }\int_{\rho ^{\ast }}^{\infty }\rho ^{s-b-1}\left[
24\pi \rho -\frac{3\mathcal{C}}{r^{2}}\right] ^{-\frac{1}{2}}d\rho  \nonumber
\\
&=&\pm \frac{2r^{2}}{3\gamma \mathcal{C}}\left( \frac{\mathcal{C}}{8\pi r^{2}%
}\right) ^{s-b}\sqrt{24\pi \rho -\frac{3\mathcal{C}}{r^{2}}}  \nonumber \\
&&\times \left. {}_{2}\mathcal{F}_{1}\left( \left[ \frac{1}{2},1+b-s\right] ,%
\left[ \frac{3}{2}\right] ,1-\frac{8\pi \rho r^{2}}{\mathcal{C}}\right)
\right\vert _{\rho ^{\ast }}^{\infty }  \nonumber \\
&=&\infty ,~~s-b\geq \frac{1}{2},  \label{TimeMRmG}
\end{eqnarray}%
where $\mathcal{C}<0$. Considering (\ref{TimeMRmG}), one can reach to the
conclusion that the potential singularity is achieved at infinity for $%
s-b\geq 1/2$, so it is not a finite-time singularity, i.e., not physical.

On the other hand, we can consider $\rho ^{a+1}$ or $\rho ^{2b}$ as a
dominant term in the integration (\ref{IntMRmG}) to show that one can have a
nonsingular universe with special constraint
\begin{eqnarray}
&&for~\ a+1 >2b:  \nonumber \\
\nonumber \\
t&=&\pm \frac{1}{\gamma \sqrt{24\pi }}\int_{\rho ^{\ast }}^{\infty }\rho
^{s-1-\frac{a+1}{2}}d\rho  \nonumber \\
&=&\pm \left. \frac{\rho ^{s-\frac{a+1}{2}}}{\gamma \sqrt{24\pi }\left( s-%
\frac{a+1}{2}\right) }\right\vert _{\rho ^{\ast }}^{\infty }=\infty ,~s-%
\frac{a}{2}\geq \frac{1}{2}, \\
\nonumber \\
\nonumber \\
&&for~\ a+1 <2b:  \nonumber \\
\nonumber \\
t&=&\pm \frac{r}{\gamma \sqrt{-3\mathcal{C}}}\int_{\rho ^{\ast }}^{\infty
}\rho ^{s-1-b}d\rho  \nonumber \\
&=&\pm \left. \frac{r\rho ^{s-b}}{\gamma \sqrt{-3\mathcal{C}}(s-b)}%
\right\vert _{\rho ^{\ast }}^{\infty }=\infty ,~~s-b\geq 0,
\end{eqnarray}%
where $\mathcal{C}<0$. This means that when $a+1>2b$ ($a+1<2b$), the time to
reach the potential singularity is infinite for $s-\frac{a}{2}\geq \frac{1}{2%
}$ ($s-b\geq 0$).\newline

\section{Nonsingular Rainbow Universe}

Here, we are going to investigate nonsingular rainbow universe by
considering some special cases of rainbow functions. The energy functions of
gravity's rainbow are motivated from different branches of the physics. The
first model comes from the hard spectra of gamma rays motivation with the
following form \cite{Amelino}
\begin{equation}
k(\varepsilon )=\frac{\exp \left( \varepsilon \right) -1}{\varepsilon },\ \
\ \ \ \ \ \ \ \ g(\varepsilon )=1.  \label{MDR1}
\end{equation}

Taking the constancy of the velocity of light into account, one can find
following relations for the rainbow functions as second model
\begin{equation}
k(\varepsilon )=g(\varepsilon )=\frac{1}{1-\varepsilon }.  \label{MDR2}
\end{equation}

The third model is motivated from loop quantum gravity and non-commutative
geometry in which rainbow functions are \cite{JacobU}
\begin{equation}
k(\varepsilon )=1,\ \ \ \ \ \ \ \ \ \ g(\varepsilon )=\sqrt{1-\varepsilon
^{n}},  \label{MDR3}
\end{equation}

Considering Eq. (\ref{averageEMR}), one can convert the above equations into
\begin{equation}
\begin{array}{ccc}
\hline\hline
Model & k(\mathcal{G}) & g(\mathcal{G}) \\ \hline\hline
first~model & \frac{\exp \left( \gamma \mathcal{G}^{\frac{\gamma -1}{\gamma }%
}\right) -1}{\gamma \mathcal{G}^{\frac{\gamma -1}{\gamma }}} & 1 \\ \hline
second~ model & \frac{1}{1-\gamma \mathcal{G}^{\frac{\gamma -1}{\gamma }}} &
\frac{1}{1-\gamma \mathcal{G}^{\frac{\gamma -1}{\gamma }}} \\ \hline
third~ model & 1 & \sqrt{1-\gamma ^{n}\mathcal{G}^{\frac{n(\gamma -1)}{%
\gamma }}} \\ \hline\hline
\end{array}
\label{MDR}
\end{equation}
where $\mathcal{G}=\rho /\rho _{P}$, and $E_{p}=c^{\prime }\rho
_{p}^{(\gamma -1)/\gamma }$ is the Planck energy versus density $\rho _{p}$.
Using the modified Friedmann equation (\ref{FE1}), on can show that
\begin{equation}
\left( H-\frac{\dot{g}(\varepsilon )}{g(\varepsilon )}\right) =\pm \frac{1}{%
k(\mathcal{G})}\sqrt{\frac{8\pi G\rho _{p}\mathcal{G}}{3}-\frac{\mathcal{C}%
m^{2}}{3r^{2}}}.  \label{FeqMDR}
\end{equation}

Considering the modified Friedmann equation (\ref{FeqMDR}) with Eq. (\ref%
{ConsMR}), following relation is obtained
\begin{equation}
\dot{\mathcal{G}}=\pm \frac{\gamma \mathcal{G}}{k(\mathcal{G})}\sqrt{24\pi
G\rho _{p}\mathcal{G}-\frac{3\mathcal{C}m^{2}}{r^{2}}},  \label{ConsVSfMDR}
\end{equation}%
in which $\dot{\mathcal{G}}=\dot{\rho}/\rho _{p}$. Now, considering the
previous discussion, we are in a position to discuss the possibility of
nonsingular universe for three cases of rainbow functions (\ref{MDR}).

\subsection{First model}

Substituting the first model of rainbow functions (\ref{MDR}) in (\ref%
{ConsVSfMDR}), on can get
\begin{equation}
\dot{\mathcal{G}}=\pm \frac{\gamma ^{2}\mathcal{G}^{\frac{2\gamma -1}{\gamma
}}}{\exp \left( \gamma \mathcal{G}^{\frac{\gamma -1}{\gamma }}\right) -1}%
\sqrt{24\pi G\rho _{p}\mathcal{G}-\frac{3\mathcal{C}m^{2}}{r^{2}}},
\label{g1}
\end{equation}

Now, we intend to elaborate the infinity of time for going from an initial
finite density $\mathcal{G}^{\ast }$\ to an infinite one in special case $%
\gamma =4/3$, (i.e., radiation). To do so, we integrate Eq. (\ref{g1})%
\begin{equation}
t=\pm \frac{9}{16}\int_{\mathcal{G}^{\ast }}^{\infty }\frac{\exp \left(
\frac{4}{3}\mathcal{G}^{\frac{1}{4}}\right) -1}{\mathcal{G}^{\frac{5}{4}}%
\sqrt{24\pi G\rho _{p}\mathcal{G}-\frac{3\mathcal{C}m^{2}}{r^{2}}}}d\mathcal{%
G},  \label{TimeMDR1}
\end{equation}%
which is not possible to obtain analytical solution, but numerical
evaluation shows that it does not converge on $\left[ \mathcal{G}^{\ast
},\infty \right) $, which leads to infinity of time to reach infinite
density. It is worthwhile to mention that this result is valid for $\gamma
>1 $. In order to make more clarification, we consider $G-$term as dominant
one in denominator which yields
\begin{eqnarray}
t &=&\pm \frac{9}{32\sqrt{6\pi G\rho _{p}}}\int_{\mathcal{G}^{\ast
}}^{\infty }\frac{\exp \left( \frac{4}{3}\mathcal{G}^{\frac{1}{4}}\right) -1%
}{\mathcal{G}^{\frac{7}{4}}}d\mathcal{G}  \nonumber \\
&=&\pm \left. \frac{\mathcal{K}_{1}-32\mathcal{G}^{\frac{3}{4}}\mathcal{E}%
\left( \frac{4}{3}\mathcal{G}^{\frac{1}{4}}\right) }{72\sqrt{6\pi G\rho _{p}}%
\mathcal{G}^{\frac{3}{4}}}\right\vert _{\mathcal{G}^{\ast }}^{\infty
}=\infty ,
\end{eqnarray}%
in which%
\begin{equation}
\mathcal{K}_{1}=3\left[ \left( 6\mathcal{G}^{\frac{1}{4}}+8\mathcal{G}^{%
\frac{1}{2}}+9\right) \exp \left( \frac{4}{3}\mathcal{G}^{\frac{1}{4}%
}\right) -9\right] ,
\end{equation}%
where $\mathcal{E}$ is the exponential integration. Obtained result
indicates that the time to reach this infinite density is infinite, so there
is no finite-time singularity which confirms the consequence of Eq. (\ref%
{TimeMDR1}).

\subsection{Second model}

Now, we are going to use the second model to obtain nonsingular universe.
Using the second form of rainbow functions (\ref{MDR}) with (\ref{ConsVSfMDR}%
), on can show that
\begin{equation}
\dot{\mathcal{G}}=\pm \gamma \left( \mathcal{G}-\gamma \mathcal{G}^{\frac{%
2\gamma -1}{\gamma }}\right) \sqrt{24\pi G\rho _{p}\mathcal{G}-\frac{3%
\mathcal{C}m^{2}}{r^{2}}},  \label{q2}
\end{equation}%
where for a special case $\gamma =4/3$, (i.e., radiation) it reduces to%
\begin{equation}
t=\pm \frac{3}{4}\int_{\mathcal{G}^{\ast }}^{\infty }\frac{d\mathcal{G}}{%
\left( \mathcal{G}-\frac{4}{3}\mathcal{G}^{\frac{5}{4}}\right) \sqrt{24\pi
G\rho _{p}\mathcal{G}-\frac{3\mathcal{C}m^{2}}{r^{2}}}},  \label{TimeMDR2}
\end{equation}%
which is not possible to obtain analytical solution. When we choose the term
with $G$\ as dominant term, the equation (\ref{TimeMDR2})\ will be%
\begin{eqnarray}
t &=&\pm \frac{3}{8\sqrt{6\pi G\rho _{p}}}\int_{\mathcal{G}^{\ast }}^{\infty
}\frac{d\mathcal{G}}{\left( \mathcal{G}^{\frac{3}{2}}-\frac{4}{3}\mathcal{G}%
^{\frac{7}{4}}\right) }  \nonumber \\
&=&\pm \frac{1}{12\sqrt{6\pi G\rho _{p}}}\left[ 3\left( \mathcal{G}^{\ast
}\right) ^{-\frac{1}{2}}\left( 3+8\left( \mathcal{G}^{\ast }\right) ^{\frac{1%
}{4}}\right) \right.  \nonumber \\
&&+\left. 32\ln \left( 1-\frac{3}{4}\left( \mathcal{G}^{\ast }\right) ^{-%
\frac{1}{4}}\right) \right] ,\ \ \ \left( \mathcal{G}^{\ast }> \left(\frac{3%
}{4}\right)^{4} \right) ,  \label{eqc2}
\end{eqnarray}%
where shows that this integration has finite value for $\mathcal{G}^{\ast
}>\left(\frac{3}{4}\right)^{4}$ and it does not converge on $(0,\left(\frac{3%
}{4}\right)^{4}]$. So, this divergency in integration is because of initial
density $\mathcal{G}^{\ast }$ and one concludes that we have big bang
singularity.

In order to obtain nonsingular universe, one may follow the previous
discussion and consider the two constants $G$ and $m$ as functions of
energy. So, one can rewrite the integration (\ref{TimeMDR2}) as
\begin{equation}
t=\pm \frac{3}{4}\int_{\mathcal{G}^{\ast }}^{\infty }\frac{d\mathcal{G}}{%
\left( \mathcal{G}-\frac{4}{3}\mathcal{G}^{\frac{5}{4}}\right) \sqrt{24\pi
\rho _{p}\mathcal{G}^{a^{\prime }+1}-\frac{3\mathcal{C}}{r^{2}}\mathcal{G}%
^{2b^{\prime }}}},  \label{AAA}
\end{equation}%
where $G(\mathcal{G})=\mathcal{G}^{a^{\prime }}$\ and $m(\mathcal{G})=%
\mathcal{G}^{b^{\prime }}$. As it was mentioned before, we may consider one
dominant term in denominator of Eq. (\ref{AAA}) which yields
\begin{eqnarray}
&&for~ a^{\prime }+1 >2b^{\prime }:  \nonumber \\
\nonumber \\
t&=&\pm \frac{3}{8\sqrt{6\pi \rho _{p}}}\int_{\mathcal{G}^{\ast }}^{\infty }%
\frac{d\mathcal{G}}{\mathcal{G}^{\frac{a^{\prime }}{2}}\left( \mathcal{G}^{%
\frac{3}{2}}-\frac{4}{3}\mathcal{G}^{\frac{7}{4}}\right) }  \nonumber \\
&=&\pm \left. \frac{3(2a^{\prime }+1)+8\mathcal{G}^{\frac{1}{4}}(a^{\prime
}+1)\mathcal{W}_{2}}{4(a^{\prime }+1)(2a^{\prime }+1)\sqrt{6\pi \rho _{p}}%
\mathcal{G}^{\frac{a^{\prime }+1}{2}}}\right\vert _{\mathcal{G}^{\ast
}}^{\infty }  \nonumber \\
&=&\infty ,~~a^{\prime }\leq -\frac{3}{2}\ \&\ \mathcal{G}^{\ast }>\left(%
\frac{3}{4}\right)^{4}
\end{eqnarray}%
\begin{eqnarray}
&&for~\ a^{\prime }+1 <2b^{\prime }:  \nonumber \\
\nonumber \\
t&=&\pm \frac{3r}{4\sqrt{-3\mathcal{C}}}\int_{\mathcal{G}^{\ast }}^{\infty }%
\frac{d\mathcal{G}}{\mathcal{G}^{b^{\prime }+1}\left( 1-\frac{4}{3}\mathcal{G%
}^{\frac{1}{4}}\right) }  \nonumber \\
&=&\pm \left. r\frac{3(1-4b^{\prime })-16\mathcal{G}^{\frac{1}{4}}b^{\prime }%
\mathcal{W}_{3}}{4b^{\prime }(4b^{\prime }-1)\sqrt{-3\mathcal{C}}\mathcal{G}%
^{b^{\prime }}}\right\vert _{\mathcal{G}^{\ast }}^{\infty } \\
&=&\infty ,~~b^{\prime }\leq -\frac{1}{4}\ \&\ \mathcal{G}^{\ast }>\left(%
\frac{3}{4}\right)^{4},  \nonumber
\end{eqnarray}%
in which%
\begin{eqnarray}
\mathcal{W}_{2} &=&_{2}\mathcal{F}_{1}\left( \left[ 1,-2a^{\prime }-1\right]
,\left[ -2a^{\prime }\right] ,\frac{4}{3}\mathcal{G}^{\frac{1}{4}}\right) ,
\\
\mathcal{W}_{3} &=&_{2}\mathcal{F}_{1}\left( \left[ 1,1-4b^{\prime }\right] ,%
\left[ 2(1-2b^{\prime })\right] ,\frac{4}{3}\mathcal{G}^{\frac{1}{4}}\right)
,
\end{eqnarray}%
where $\mathcal{C}<0$ and the initial density $\mathcal{G}^{\ast }$ should
be larger than $(3/4)^{4}$. The above equations mean that when $a^{\prime
}+1>2b^{\prime }$ ($a^{\prime }+1<2b^{\prime }$), the time to reach the
potential singularity is infinite for $a^{\prime }\leq -\frac{3}{2} $\ ($%
b^{\prime }\leq -\frac{1}{4}$).

\subsection{Third model}

Substituting the third model of rainbow functions (\ref{MDR}) in (\ref%
{ConsVSfMDR}), it is easy to show that
\begin{equation}
\dot{\mathcal{G}}=\pm \gamma \mathcal{G}\sqrt{24\pi G\rho _{p}\mathcal{G}-%
\frac{3\mathcal{C}m^{2}}{r^{2}}},  \label{g3}
\end{equation}%
which leads to the following integration
\begin{eqnarray}
t &=&\pm \frac{1}{\gamma }\int_{\mathcal{G}^{\ast }}^{\infty }\frac{d%
\mathcal{G}}{\mathcal{G}\sqrt{24\pi G\rho _{p}\mathcal{G}-\frac{3\mathcal{C}%
m^{2}}{r^{2}}}}  \nonumber \\
&=&\pm \frac{2r}{\gamma m\sqrt{-3\mathcal{C}}}\sinh ^{-1}\left( \frac{m}{2r}%
\sqrt{-\frac{\mathcal{C}}{2\pi G\mathcal{G}^{\ast }}}\right) ,
\label{TimeMDR3}
\end{eqnarray}%
where $\mathcal{C}<0$ and shows that we have singular universe. It is
worthwhile to mention that the integration (\ref{TimeMDR3}) is like Eq. (\ref%
{TM}) and one may note that the rainbow function, $g(\varepsilon )$, does
not affect the singularity of universe. But here, there is another story
because of rainbow functions. Indeed, we can follow the previous discussion
to consider the two constants $G$ and $m$ as functions of energy in order to
obtain nonsingular universe. So, the integration (\ref{TimeMDR3}) will
convert to
\begin{equation}
t=\pm \frac{1}{\gamma }\int_{\mathcal{G}^{\ast }}^{\infty }\frac{d\mathcal{G}%
}{\mathcal{G}\sqrt{24\pi \rho _{p}\mathcal{G}^{a^{\prime \prime }+1}-\frac{3%
\mathcal{C}}{r^{2}}\mathcal{G}^{2b^{\prime \prime }}}},
\end{equation}%
where $G(\mathcal{G})=\mathcal{G}^{a^{\prime \prime }}$\ and $m(\mathcal{G})=%
\mathcal{G}^{b^{\prime \prime }}$. As it has mentioned before, one may
consider a dominant term to solve the integration
\begin{eqnarray}
&&for~ a^{\prime \prime }+1>2b^{\prime \prime }:  \nonumber \\
\nonumber \\
t&=&\pm \frac{1}{2\gamma \sqrt{6\pi \rho _{p}}}\int_{\mathcal{G}^{\ast
}}^{\infty } \frac{d\mathcal{G}}{\mathcal{G}^{\frac{a^{\prime \prime }+3}{2}}%
}  \nonumber \\
&=&\pm \left. \frac{1}{\gamma (a^{\prime \prime }+1)\sqrt{6\pi \rho _{p}}%
\mathcal{G}^{\frac{a^{\prime \prime }+1}{2}}}\right\vert _{\mathcal{G}^{\ast
}}^{\infty }=\infty ,~~a^{\prime \prime }\leq -1, \\
\nonumber \\
\nonumber \\
&&for~ a^{\prime \prime }+1<2b^{\prime \prime }:  \nonumber \\
\nonumber \\
t&=&\pm \frac{r}{\gamma \sqrt{-3\mathcal{C}}}\int_{\mathcal{G}^{\ast
}}^{\infty }\frac{d\mathcal{G}}{\mathcal{G}^{b^{\prime \prime }+1}}
\nonumber \\
&=&\pm \left. \frac{r}{\gamma b^{\prime \prime }\sqrt{-3\mathcal{C}}\mathcal{%
G}^{b^{\prime \prime }}}\right\vert _{\mathcal{G}^{\ast }}^{\infty }=\infty
,~~b^{\prime \prime }\leq 0,
\end{eqnarray}%
where $\mathcal{C}<0$. This means that when $a^{\prime \prime }+1>2b^{\prime
\prime }$ ($a^{\prime \prime }+1<2b^{\prime \prime }$), the time to reach
the potential singularity is infinite for $a^{\prime \prime }\leq -1$ ($%
b^{\prime \prime }\leq 0$) which is true for all values of $\gamma $
including $\gamma =4/3$.\newline

\subsection{Density of states}

Our final study is devoted to the possible divergency of density of states
at the Planck scale \cite{cosmosLing,LingWu}. Employing the modified
dispersion relation (\ref{DeforEM}), the density of states could be obtained
as
\begin{equation}
a(E)dE\simeq p^{2}dp=\frac{k(\varepsilon )^{3}}{g(\varepsilon )^{3}}\left[
1+E\left( \frac{k(\varepsilon )^{\prime }}{k(\varepsilon )}-\frac{%
g(\varepsilon )^{\prime }}{g(\varepsilon )}\right) \right] E^{2}dE,
\end{equation}%
in which by remembering the fact that energy cannot be larger than Planck
energy, the density of states yield to a finite value for all rainbow
function models (Eqs. (\ref{MDR1})--(\ref{MDR3}))
\begin{equation}
a(E)\simeq \left\{
\begin{array}{cc}
\frac{E^{2}}{\varepsilon ^{2}}exp(\varepsilon) \left[ exp(\varepsilon)-1%
\right] ^{2}, & first~model, \\
&  \\
E^{2}, & second\ model, \\
&  \\
\frac{E^{2}\left[ (n-2)\varepsilon ^{n}+2\right] }{2(1-\varepsilon
^{n})^{5/2}}, & third~model,%
\end{array}%
\right. ,
\end{equation}%
which show that the density of states have regular behavior without any
divergency (note: $\varepsilon<1$).\newline

\section{Closing Remarks}

Motivated by the high energy regime at the early universe, developing
Einstein gravity has been applied in the context of cosmology. FRW cosmology
in the presence of massive gravity and massive gravity's rainbow have been
separately investigated. First, the massive gravity modification was
investigated and it was shown that generalization to massive gravity does
not remove the big bang singularity. Then, the generalization to gravity's
rainbow was imposed for two different cases; in one case, the constants were
considered independent of energy while in the other case the energy
dependency of constants was taken into account. It was pointed out that in
order to remove the big bang singularity in an energy dependent spacetime,
certain conditions are required to be satisfied.

Using the method which was inscribed in \cite{Amelino} (choosing the
suitable rainbow functions), it was possible to study the effects of rainbow
functions on FRW-like cosmology. It was shown that Friedmann equations were
modified in the presence of massive gravity's rainbow which lead to the
absence of big bang singularity. Such property was derived for large varying
range of equation of state parameter, $\gamma >4/3$. The absence of
singularity was shown by using the analysis in \cite%
{cosmosAwad,Hendicosmology}. It was found that the universe takes infinite
time to reach $\rho \rightarrow \infty $ from an initial finite value of $%
\rho$. Then, we have investigated two other models of rainbow functions (see
Eqs. (\ref{MDR2}) and (\ref{MDR3})) and we found that the universe will be
singular in these two cases. In order to obtain nonsingular universe for
these two models, we had to consider two constants $G$ and $m$ as functions
of energy. Finally, the possibility of divergency of density of state at the
Planck scale was investigated for three models of rainbow functions. It was
pointed out that considering the energy conditions of gravity's rainbow, the
density of state does not diverge and a possible resolution regarding the
big bang singularity is obtained.

Here, we have provided a preliminary to study the effects of massive
gravitons on different stages of the universe's evolution, especially
gravitational waves which were produced in these stages. It is interesting
to study the effects of gravity's rainbow and massive gravity on the
inflation mechanism and the age of different eras in the standard cosmology.
In addition, it is worthwhile to see how these two generalizations could
address the old cosmological constant problem and accelerating expansion of
the universe.\newline


\textbf{Acknowledgements}

We would like to thank the referee for his/her insightful comments
which lead to significant improvement in the paper. We also wish
to thank Shiraz University Research Council. This work has been
supported financially by the Research Institute for Astronomy and
Astrophysics of Maragha, Iran.

\end{document}